\documentclass[12pt]{article}
\usepackage{amsmath, amssymb}
\usepackage{geometry}
\usepackage{times}
\usepackage{url}
\usepackage{amsmath}
\usepackage{graphicx}
\usepackage[colorlinks=true, allcolors=blue]{hyperref}

\title{\textbf{Pink Noise in Economic Time Series from Synchronization and Amplitude Demodulation}}
\author{
    Masahiro Morikawa$^{\ast}$, Yokoh Morikawa$^{\dagger}$, Akika Nakamichi$^{\ddagger}$ \\
    \\
    $^{\ast}$RIKEN, Wako Saitama, Japan 
    \thanks{hiro@phys.ocha.ac.jp} \\
    $^{\dagger}$California State University Long Beach, USA \\
    $^{\ddagger}$Kyoto Sangyo University, Japan
}
\date{}

\begin{document}
\maketitle

\abstract{
Pink noise, characterized by a power spectral density $S(\omega)\propto\omega^{\beta}$ with $\beta\simeq -1$, appears in economic indices as well as in many natural systems.  We summarize a unified mesoscopic interpretation in which pink spectra arise from repeated synchronization, amplitude modulation, and demodulation. 
In economic time series, we identify two kinds of pink-noise behavior:
one that appears in the raw data (property A), and another that appears
only after detrending and demodulation (property B).
A stochastic Kuramoto model provides a minimal dynamical model of repeated synchronization and desynchronization among many economic circulations.  It produces approximate $1/f$ spectra over a broad coupling--system-size domain and gives variance--mean scaling, Taylor's law.  The same amplitude-modulation/demodulation mechanism also gives a compact explanation of pink spectra in music, earthquakes, variable stars, solar flares, and black-hole accretion systems.  Pink noise is therefore interpreted not merely as a statistical regularity, but as a diagnostic of slowly modulated collective coherence in complex flow systems.
} 

\section{Introduction} 
\label{sec:intro}
Pink noise, or $1/f$ fluctuation, is usually defined by a power spectral density (PSD)
\begin{equation}
 P(\omega)=A\omega^{\beta},\qquad \beta\simeq -1\pm0.3,
 \label{eq:pink}
\end{equation}
over a finite frequency \(\omega\) range.  Since the first observation of low-frequency excess noise in vacuum tubes by Johnson~\cite{Johnson1925}, $1/f$ spectra have been reported in electrical currents, biological rhythms, climate records, earthquakes, music, economic activity, solar flares, stellar variability, and accretion systems~\cite{WardGreenwood2007,Press1978}.  In finance, related long-memory and volatility-clustering phenomena are among the standard stylized facts of markets~\cite{Mandelbrot1963,Engle1982,Cont2001,MantegnaStanley2000}.  Many mechanisms have been proposed, including superposed relaxation processes~\cite{Machlup1954}, carrier-number fluctuations~\cite{McWhorter1957}, mobility fluctuations~\cite{Hooge1969}, fractal time series~\cite{MandelbrotVanNess1968,Mandelbrot1982}, self-organized criticality~\cite{BakTangWiesenfeld1987,Bak1996}, and fluctuating transport coefficients in accretion disks~\cite{Lyubarskii1997,UttleyMcHardyVaughan2005}.  These theories are useful in their own domains, but the breadth of $1/f$ phenomenology suggests that an additional, more kinematic route may also exist.

The aim of this article is to describe such a route, with economic indices as the central application.  The proposed mechanism, developed in our previous amplitude-modulation studies~\cite{MorikawaNakamichi2023SciRep}, has three elements: (i) many interacting oscillatory or circulatory units, (ii) partial synchronization that produces beats and slow amplitude modulation (AM), and (iii) nonlinear observation, dissipation, thresholding, transaction, rupture, radiation, or other demodulation (DM) that converts the slow envelope into an observable low-frequency signal.  In this view, pink noise does not necessarily require fine tuning to a single critical point.  It can arise over a broad critical domain in which many collective modes repeatedly synchronize and desynchronize.

Economic systems are especially suitable for this viewpoint.  They contain many loops: production--income--consumption loops, inventory loops, credit loops, inter-firm supply loops, expectation--price loops, and international trade loops.  Modern economic-network studies make this statement concrete: firms, banks, and households can be represented as heterogeneous nodes connected by directed flows of goods, services, credit, and money, and such flows can be analyzed as large-scale weighted directed networks~\cite{Fujiwara2024}.  In particular, Helmholtz--Hodge decomposition has been used to separate economic-network flows into potential, or gradient, components and divergence-free circulatory components, for example in Bitcoin money-flow networks and firm-account money-flow networks~\cite{FujiwaraIslam2020,Fujiwara2021EPJDS}.  These loops do not oscillate independently.  They entrain through prices, credit conditions, inventories, transportation constraints, policy, and expectations.  The central question is therefore not only whether an economic index has a $1/f$ spectrum, but what is being synchronized and what operation demodulates the hidden envelope.
\section{Type A, Type B, Type AB, and Type O pink noise in economic data}
\label{sec:typeABO}
In this paper we classify pink noise in economic time series by separating two
different questions.  The first is whether the original observable itself has a
pink spectrum.  The second is whether the magnitude of its detrended
fluctuations has a pink spectrum.  Let $x(t)$ denote an economic observable,
and let
\begin{equation}
    \delta x(t) = x(t)-\bar{x}_{\tau}(t)
\end{equation}
be the residual after removing a local trend $\bar{x}_{\tau}(t)$ over a
timescale $\tau$.  We then define a fluctuation-amplitude proxy
\begin{equation}
a(t)=|\delta x(t)|,\quad
a(t)=[\delta x(t)]^2,\quad \text{or}\quad
a(t)=\max[\delta x(t),0],
\end{equation}
or, when appropriate, the absolute return or absolute increment.  The power
spectra of $x(t)$ and $a(t)$ are denoted by $P_x(\omega)$ and $P_a(\omega)$,
respectively.  A signal is called pink when, over a finite but statistically
meaningful frequency range,
\begin{equation}
    P(\omega) \propto \omega^{\beta}, \qquad \beta \simeq -1 .
\end{equation}

\begin{figure}[t]
\centering
\includegraphics[width=1\textwidth]{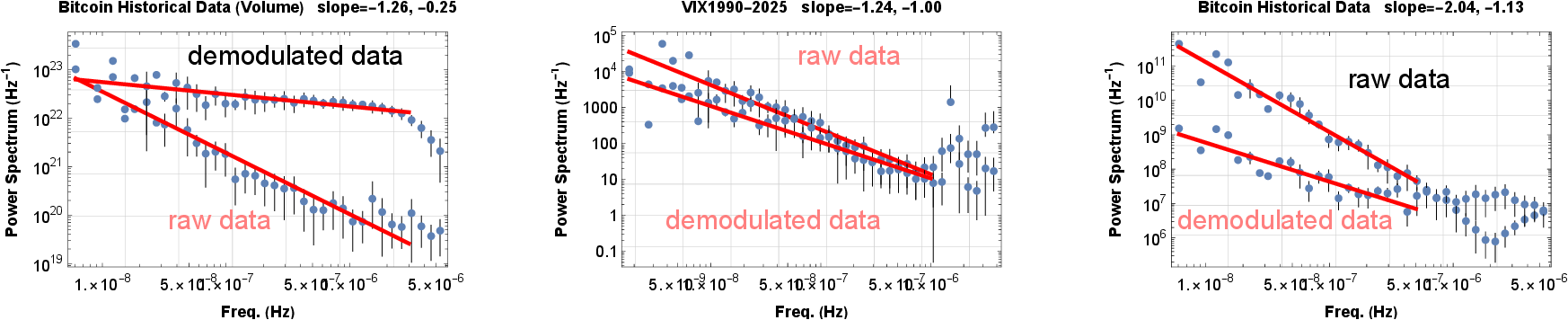}
\caption{Some empirical classes of economic pink noise in PSD. 
(left) Raw Bitcoin volume data from Yahoo Finance  \cite{YahooFinanceBTC} exhibit pink noise (PSD power index is -1.26), but the detrended squared data is almost white (-0.25). 
Thus, the data is classified as Type A. 
(center) VIX index historical data \cite{YahooFinanceVIX} exhibit pink noise (-1.24), and  the detrended squared data also exhibit pink noise (-1.00). 
Thus, the data is classified as Type AB. 
The classification should therefore be regarded as operational rather than absolute. In particular, the distinction between Type A and Type AB can depend on the degree of intrinsic demodulation already present in the observational process, as well as on the chosen explicit demodulation method.
(right) Raw Bitcoin price historical data \cite{YahooFinanceBTC} exhibit red noise (-2.04) but the detrended squared data is pink (-1.13). 
Thus, the data is classified as Type B. 
\newline 
There are many economic indices that do not exhibit pink noise. For example, the Federal Funds Rate \cite{FRED_FEDFUNDS} is strongly red (-2.99), and the detrended squared data is white (0.03).
\newline 
Although the coefficient of determination is not explicitly shown, most fitted values have $R^2>0.9$ throughout this paper. 
We did not apply window functions in the displayed analysis because tests with standard windows did not materially change the broadband PSD slopes.
In this paper, we consistently use the cubic S-spline method for detrending.  
}
\label{fig:typeab}
\end{figure}

This gives a natural two-axis classification.  Type A pink noise is the case in
which the raw observable already has a pink spectrum,
    \(P_x(\omega) \propto \omega^{-1}\),
whereas the detrended fluctuation amplitude does not show a clear pink
spectrum.  
Type B pink noise is the opposite case: the raw observable is not
pink, but the detrended amplitude is,
    \(P_a(\omega) \propto \omega^{-1}\).
We further distinguish two additional cases.  Type AB denotes observables for
which both the raw signal and the fluctuation amplitude show pink spectra,
    \(P_x(\omega) \propto \omega^{-1}, 
    P_a(\omega) \propto \omega^{-1}\).
Finally, Type O denotes observables for which neither the raw signal nor the
detrended amplitude shows robust pink scaling.  The four cases can be
summarized as
\begin{center}
\begin{tabular}{c|cc}
\hline
Type & raw signal $x(t)$ & detrended amplitude $a(t)$ \\
\hline
A  & pink     & not pink \\
B  & not pink & pink \\
AB & pink     & pink \\
O  & not pink & not pink \\
\hline
\end{tabular}
\end{center}
We show typical examples of these classes in Fig.\ref{fig:typeab}. 

The two axes, raw-signal pinkness and demodulated-amplitude pinkness,
represent two different layers of temporal organization.
Pinkness in the raw
signal indicates long-memory organization in the level, volume, or flow itself.
Pinkness in the detrended amplitude indicates long-memory organization in the
strength of fluctuations, i.e. volatility clustering or amplitude modulation.
Thus Type A is naturally associated with extensive or flow-like quantities,
such as production, sales volume, transaction volume, or other activity
measures whose raw magnitudes are generated by many coupled economic
circulations.  Type B is naturally associated with intensive or price-like
quantities, such as returns, exchange-rate changes, interest-rate changes, or
other quantities whose levels may not be pink but whose fluctuation amplitudes
are strongly clustered.  Type AB corresponds to systems in which both the
activity level and the fluctuation amplitude retain long memory.  Type O is the
null class, in which neither layer shows a robust pink signature. 

This refinement is useful because Type A and Type B pink noise should not be
interpreted as the same phenomenon.  Type A suggests that the observable itself
is a direct projection of coupled circulation modes.  In contrast, Type B
suggests that the observable is closer to a response variable whose signed
fluctuations may decorrelate rapidly, while their amplitude is modulated by a
slow collective state.  In economic language, Type A is closer to persistent
flow, whereas Type B is closer to persistent risk, volatility, or market
tension.  Type AB is especially important because it indicates that both
mechanisms coexist: the economic flow has long memory, and the intensity of its
deviations also has long memory.  In such a case, a single scalar PSD slope is
insufficient; one must specify whether the pink spectrum appears in the raw
level, in the detrended amplitude, or in both.

Although this notation is unrelated to the blood-group system, a useful physiological analogy is biological circulation.  Heartbeat intervals are a classical example of approximately $1/f$-like physiological variability~\cite{KobayashiMusha1982}.  More recently, cortical tissue oxygenation was shown to have clear $1/f$-like spectra, and stochastic heterogeneity in red-blood-cell spacing and flow was identified as one origin of these fluctuations~\cite{ZhangGheresDrew2021}.  This analogy should not be read as a direct claim that every blood-flow variable is pink; rather, it supports the broader idea that transport networks driven by many interacting pulses, stalls, and local circulations can naturally generate pink-like fluctuations. 

\section{Amplitude modulation and demodulation}
\label{sec:amdm}
We propose that one origin of pink noise is the amplitude modulation (AM) and demodulation (DM). 
The simplest mechanism of AM/DM is the beat formation. 
For two nearby frequencies (\(0<\lambda \ll \omega\)),
\begin{equation}
 \sin[(\omega+\lambda)t]+\sin[(\omega-\lambda)t]
 =2\cos(\lambda t)\sin(\omega t)
 \equiv A(t)\sin(\omega t),
 \label{eq:beat}
\end{equation}
where the slowly varying envelope is $A(t)=2\cos(\lambda t)$.  The envelope is not necessarily visible in the signed signal, but it appears after an even nonlinear operation. For example,
\begin{equation}
 [A(t)\sin(\omega t)]^2
 =\frac{1}{2}A^2(t)-\frac{1}{2}A^2(t)\cos(2\omega t).
 \label{eq:square}
\end{equation}
The first term is a low-frequency envelope contribution.  Absolute values and rectification work similarly.  By contrast, odd powers keep the signed carrier more strongly and are less efficient envelope extractors.  Equation~(\ref{eq:square}) is therefore the minimal mathematical reason why Type B spectra appear in squared, absolute, or rectified residuals.

A single beat produces a characteristic frequency, not a scale-free spectrum.  To obtain pink noise, one needs a broad hierarchy of beat frequencies \cite{MorikawaNakamichi2023SciRep}.  
For example, suppose that characteristic frequencies for the oscillators are generated by an
exponential mapping, \(\omega=e^{-\lambda t}\), where \(t\) is uniformly
distributed. Then \(Q(\omega)d\omega=Q(t)dt\) yields
\begin{equation}
Q(\omega)  =Q(t)|d \omega / d t|^{-1}=q(\lambda \omega)^{-1} \propto \omega^{-1},
 \label{eq:envelope}
\end{equation}
for the frequency distribution $Q(\omega)$. 
Then the frequency-difference $\Delta \omega$ distribution also becomes inversely proportional to $\Delta \omega$, and yields pink noise. 
This is only a minimal illustrative construction; many hierarchical or
multiplicative mechanisms can generate similar broad distributions of
beat frequencies.
A power-law frequency distribution also yields a similar distribution of $\Delta \omega$ with power slightly modified from $-1$. 
Such a distribution can arise from multiplicative accumulation of time scales, hierarchical organization, or repeated synchronization among clusters.  The precise microscopic origin can differ from system to system, but the AM/DM transformation is common.

This point reverses the usual explanatory direction.  Pink noise is often explained by fractality, criticality, or fluctuating transport coefficients.  Here, once a broad AM/DM-generated pink spectrum exists, it can itself generate apparent fractal scaling, critical-like avalanches, and fluctuating effective coefficients after coarse graining.  Thus, some observations usually taken as primitive explanations may instead be secondary descriptions of a modulated collective envelope.

\section{Stochastic Kuramoto model for economic synchronization}
\label{sec:skm}
A natural minimal model of phase synchronization is the Kuramoto model~\cite{Kuramoto1984,Strogatz2000,Acebron2005},
\begin{equation}
 \frac{d\theta_i}{dt}
 =\omega_i+\frac{K}{N}\sum_{j=1}^{N}\sin(\theta_j-\theta_i),
 \label{eq:km}
\end{equation}
where $\theta_i$ is the phase of the $i$th oscillator, $\omega_i$ its natural frequency, $K$ the coupling strength, and $N$ the number of units.  In the economic interpretation, the oscillators are not literal harmonic oscillators.  They represent recurrent loops of production, inventory, credit, trade, pricing, or expectation.  Their phases indicate positions along these loops.

For scalar data $x(t)$, a phase can be associated with the analytic signal
\begin{equation}
 G(t)=x(t)+\frac{i}{\pi}{\cal P}\int_{-\infty}^{\infty}\frac{x(t')}{t-t'}dt'
      =r(t)e^{i\theta(t)},
 \label{eq:hilbert}
\end{equation}
where ${\cal P}$ denotes the Cauchy principal value.  This Hilbert-transform construction motivates the use of phase dynamics even when the original observation is a single real-valued time series.

The original Kuramoto model, however, is too static for economics.  Once phase locking occurs, the system remains locked unless parameters change.  Real economies repeatedly synchronize, desynchronize, crash locally, reorganize, and synchronize again.  We therefore use a stochastic Kuramoto model (SKM),
\begin{equation}
 \frac{d\theta_i}{dt}
 =\omega_i+\frac{K}{N}\sum_{j=1}^{N}\sin(\theta_j-\theta_i)+\xi_i(t).
 \label{eq:skm1}
\end{equation}
The associated collective order parameter is
\begin{equation}
 R(t)e^{i\Psi(t)}=\frac{1}{N}\sum_{j=1}^{N}e^{i\theta_j(t)}.
 \label{eq:order}
\end{equation}
The variable $R(t)$ measures coherence.  Its fluctuation, or a demodulated signal constructed from the synchronized output, is the natural candidate for the observed pink component.

\begin{figure}[t]
\centering
\includegraphics[width=1\textwidth]{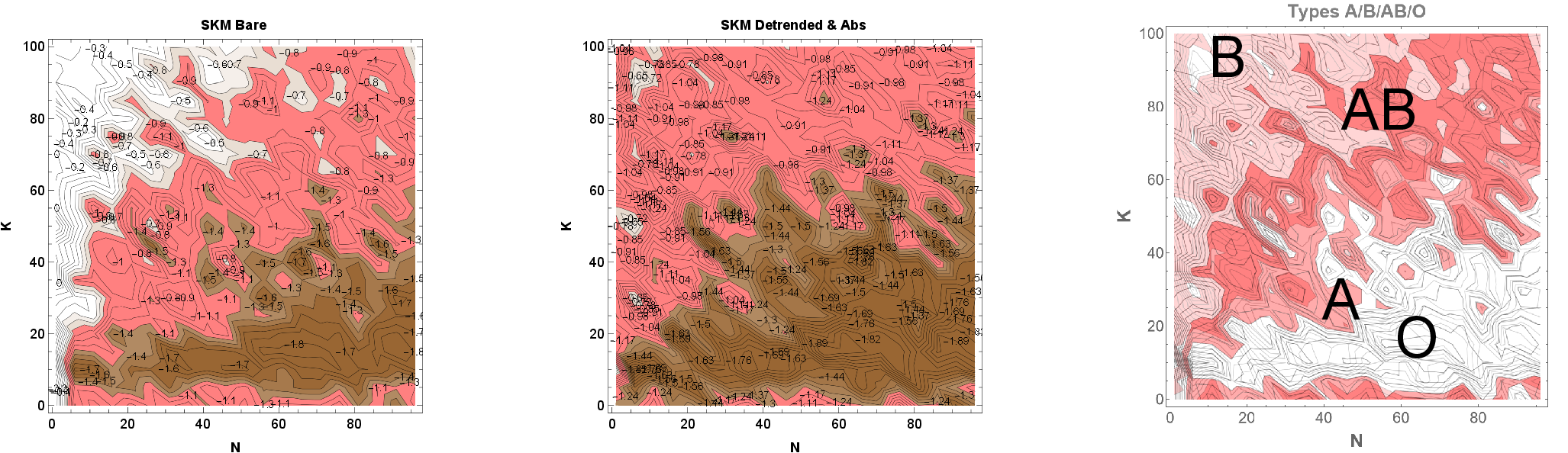}
\caption{Stochastic Kuramoto dynamics as a minimal model of repeated synchronization and desynchronization. PSD power indices are calculated for various parameters $N$ and $K$ for the order parameter. Pink noise appears in a broad parameter domain.
We mark the pink-noise regions (\(-1.25<\mathrm{index}<-0.75\)) for the raw data (left), and for the absolute value of the detrended data (center). 
The whiter and brownish regions represent shallower, white-like spectra (\(-0.75<\mathrm{index}\)) and steeper, brown-like spectra (\(\mathrm{index}<-1.25\)) respectively. 
(right) The overlap of the two pink regions defines Type AB, while the non-overlap
regions define Types A and B; the remaining region corresponds to Type O. 
\newline
The frequencies {$\omega_i$} in Eq.(\ref{eq:skm1}) are chosen randomly around $2\pi\times100$ with a width of $1$\%. 
In addition, each \(\theta_i\) is randomly reset at random time intervals with mean interval $0.05$.}
\label{fig:skm}
\end{figure}

Numerical experiments show that the SKM can produce approximate $1/f$ spectra with slopes close to $-1$ for modest $N$, finite frequency dispersion, and stochastic forcing.  More importantly, the pink behavior appears over a broad region of the $(N,K)$ parameter plane.  This suggests a critical domain rather than a critical point.
The domain structure becomes apparent in Fig.\ref{fig:skm}.    
In economic language, many different levels of coupling and diversity can yield slowly modulated coherence; no unique critical value of the coupling is required.

\begin{figure}[t]
\centering
\includegraphics[width=0.75\textwidth]{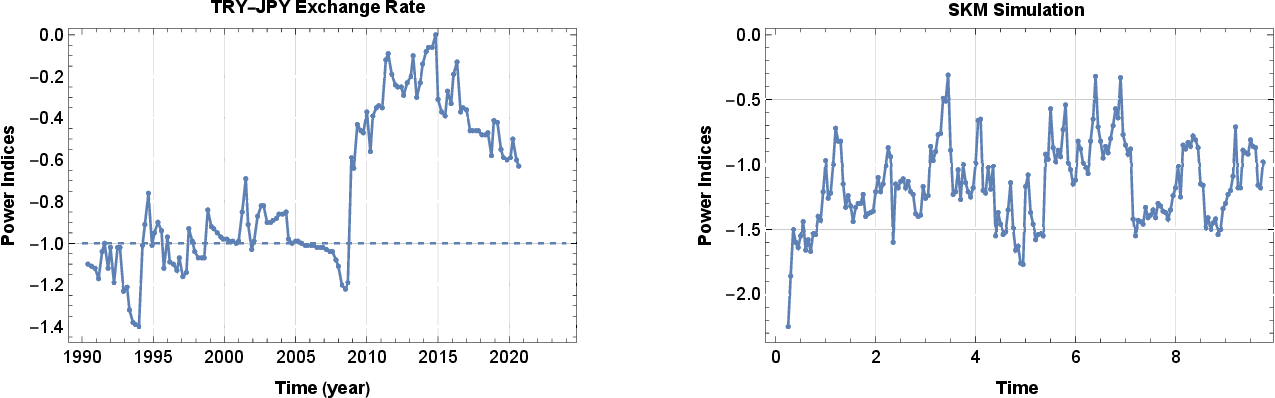}
\caption{ 
Pink-noise diagnostics in economics. The PSD exponent measures coherence in the economic indices. 
\newline
Moving-window estimates of the PSD exponent $\beta(t)$ for
(left) TRY-JPY exchange rate historical data from Investing.com \cite{InvestingTRYJPY}, and (right) typical SKM calculation. 
The abrupt whitening around 2008--2009 is plausibly associated with the Lehman Brothers bankruptcy and the subsequent global financial crisis.
Similar whitening-and-recovery episodes occur at multiple scales in
various economic time series.
}
\label{fig:piecewise}
\end{figure}

A moving-window estimate of the PSD exponent $\beta(t)$ can then be interpreted as a coherence index.  During normal periods the economy may remain near a pink coherent state.  Around crises, local crashes, or regime shifts, the exponent can deviate from unity.  The return toward $\beta\simeq -1$ may indicate reorganization into a new synchronized configuration. Typical examples are shown in Fig.\ref{fig:piecewise}.

In financial and macroeconomic systems, some observables are already demodulated at the measurement level. Trading volume, turnover, volatility, spreads, default counts, and adjustment intensities do not primarily record the signed phase of an economic cycle, but rather its activity envelope or stress amplitude. These variables can therefore exhibit Type A or Type AB pink noise even without an explicit absolute-value or detrending operation. The demodulation is, however, generally imperfect and asymmetric, as reflected in gain/loss asymmetry and the leverage effect in financial returns.

This gives the Type A/B/AB/O classification an observational meaning:
Type A and Type AB variables are often already envelope-like observables,
whereas Type B variables retain signed cyclic components and require
explicit demodulation before their slow collective envelope becomes visible.

The distinction between Type A/AB and Type B may be closely related to the distinction between extensive and intensive observables. Extensive variables add up the activity of many economic agents and therefore tend to measure the activity envelope. They are partially demodulated already at the observational level. Intensive variables tend to retain signed cyclic components. Their pink-noise structure may therefore become visible only after operational detrending and demodulation.

This viewpoint also suggests that the distinction between Type A and Type AB may not always be sharp. In many real systems, the observational process itself acts as an intrinsic but imperfect demodulator. As a result, the raw observable may already contain part of the slowly varying envelope, while residual carrier-like fluctuations with hidden information remain. Different operational demodulators can then reveal different aspects of this hidden structure. For example, absolute-value demodulation may preserve long-memory activity clustering, whereas squared signals can become dominated by intermittent burst amplitudes and appear whiter. In this sense, Type A, AB, and B may represent different degrees and forms of intrinsic demodulation of an underlying amplitude-modulated process rather than fundamentally distinct origins of pink noise.

\section{Taylor's law, volatility clustering, and apparent economic cycles}
\label{sec:taylor}
Economic Type B pink noise is closely related to volatility clustering, a central stylized fact of financial time series~\cite{Engle1982,Cont2001}.  When the residual $\delta x(t)$ is demodulated to $y(t)=\{\delta x(t)\}^2$ or $|\delta x(t)|$, the result is a volatility-like positive process.  A common scaling law for such processes is Taylor's law~\cite{Taylor1961,Eisler2008},
\begin{equation}
 {\rm Var}[y] \propto \langle y\rangle^\alpha,
 \qquad \alpha\simeq 2.
 \label{eq:taylor}
\end{equation}
In astrophysics the analogous statement is the rms--flux relation: the rms variability is approximately proportional to the mean flux~\cite{UttleyMcHardyVaughan2005}.  In both cases the variance is not a fixed noise level.  It is a fluctuating collective variable, consistent with multiplicative or amplitude-modulated dynamics.

\begin{figure}[t]
\centering
\includegraphics[width=0.75\textwidth]{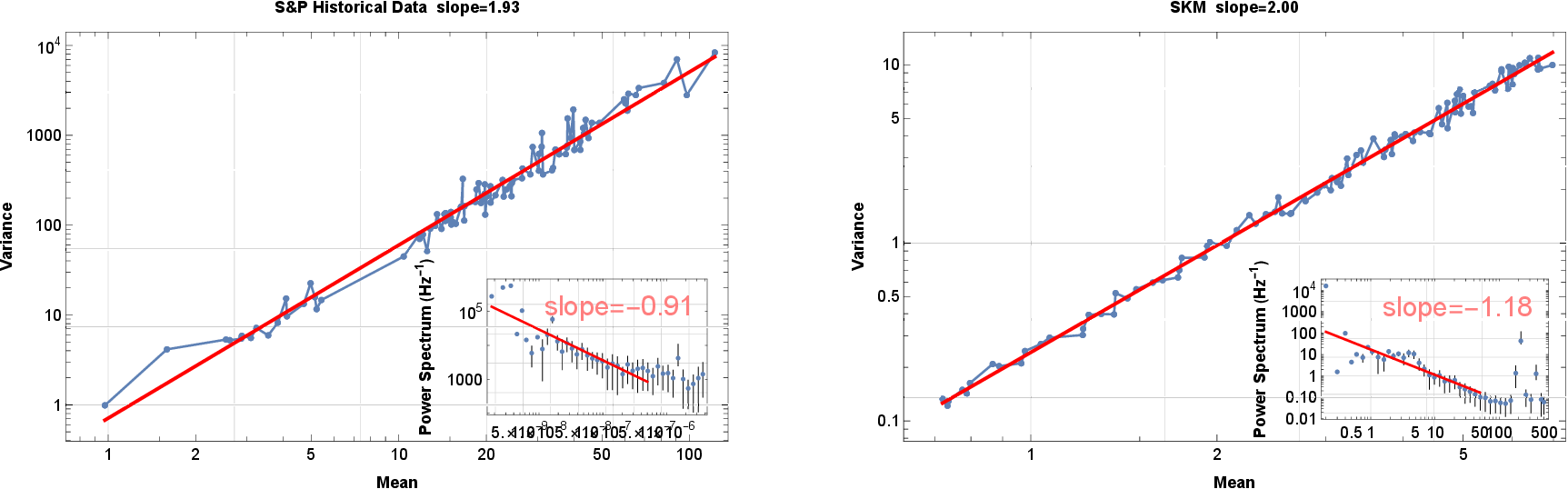}
\caption{ 
Pink-noise diagnostics in economics.  Taylor's law measures amplitude fluctuation.
Taylor's law for
(left) S\&P 500  historical data \cite{YahooFinanceGSPC},
and (right) a typical SKM calculation.
}
\label{fig:taylor}
\end{figure}

In representative financial data, detrended and absolute-valued stock-index fluctuations give variance--mean slopes close to $2$. SKM simulations also yield slopes near $1.9$ over finite ranges. 
They are displayed in Fig.\ref{fig:taylor}. 
The finite range is important.  The simplest phase-only SKM captures the emergence of pink volatility and approximate Taylor scaling, but it does not fully determine broad distribution functions or extreme tails.  A more complete model should include amplitude or participation dynamics in addition to phase dynamics.  One possible extension is
\begin{equation}
 z_i(t)=\rho_i(t)e^{i\theta_i(t)}
       =\exp[-\lambda_i(t)+i\theta_i(t)],
 \label{eq:complexphase}
\end{equation}
where $\lambda_i(t)$ represents growth, damping, liquidity, or effective participation weight.  Such amplitude--phase dynamics may be required to connect pink spectra, Taylor's law, lognormal variability, and power-law tails in a single framework.

The same framework gives a cautionary interpretation of economic cycles.  If the PSD is close to $1/f$ but not exactly scale-free over all frequencies, the autocorrelation function may show quasi-periodic peaks.  Some apparent Kitchin, Juglar, Kuznets, or Kondratieff-like periods may therefore be artificial periods produced by finite-band deviations from pink noise.  This does not mean that business-cycle mechanisms are absent.  Rather, any claimed clock-like period should be tested against a null model of modulated pink noise generated by synchronization drift.

\section{Universality beyond economics}
\label{sec:beyond}
\begin{figure}[t]
\centering
\includegraphics[width=1\textwidth]{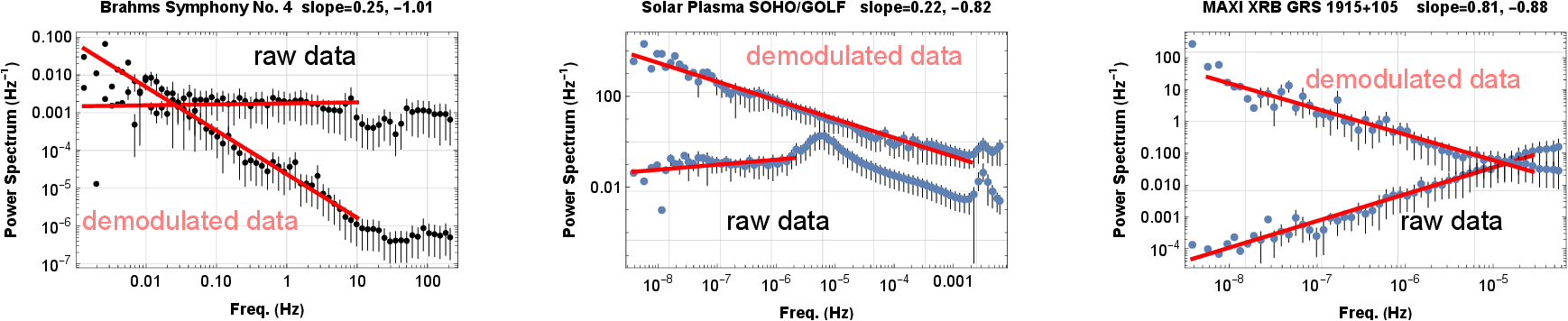}
\caption{Universality of the AM/DM mechanism. 
The synchronized units and demodulators differ by system, but the envelope-extraction logic is common.
Type B pink noise examples are taken from musical audio Brahms Symphony No. 4 \cite{InternetArchiveBrahms4} (left), solar plasma speed data \cite{SOHO_GOLF}  (center), and X-ray fluctuations from the black-hole X-ray binary GRS 1915+105 \cite{MAXI_GRS1915} (right). 
Although the PSDs of the raw data are mostly white or shallow
(shallower line in each graph), the detrended squared data exhibit
pink-noise-like spectra.
The vertical axes are slightly shifted for visibility. 
}
\label{fig:beyond}
\end{figure}

The AM/DM scenario for pink noise is not restricted to economics.  It applies whenever oscillatory or circulatory units synchronize partially and an observation process extracts an amplitude envelope. 
Several Type B examples are shown in Fig. \ref{fig:beyond}. 
Although their raw data are mostly white, the detrended squared data often exhibit pink noise. 

In music, the original waveform is an intensive oscillatory signal.  Its squared or rectified amplitude exhibits Type B pink noise.  Ensemble performance naturally supplies synchronization: multiple instruments or voices create unison, beat, 
and envelope coherence.  The observation of pink spectra in demodulated music signals is therefore structurally similar to Type B pink noise in financial indices.  A recent detailed application to music and related natural signals argues that dynamic synchronization and resonance generate amplitude modulation, while squaring or rectification extracts the low-frequency envelope~\cite{NakamichiUesakaMorikawa2026Entropy}.

In earthquakes, the energy-release sequence is a positive extensive observable and often behaves as Type A pink noise.  A complementary Type B interpretation is also possible.  Earth's free oscillations provide a set of eigenmodes~\cite{NakamichiMatsuiMorikawa2024}, and rupture or fault disruption acts as intrinsic demodulation.  A superposition of mechanical modes can therefore become a pink energy-release process after nonlinear rupture.  This connects seismic pink noise to resonance and demodulation rather than only to avalanche criticality.

Variable stars and solar flares provide astrophysical examples.  Coupled convective cells, magnetic flux tubes, or flare-producing regions can synchronize partially.  Radiation energy, magnetic reconnection, and thermal emission then demodulate the internal motion.  Coupled Lorenz-type convective models and the resonance of solar five-minute oscillation analyses illustrate how local oscillations can create pink radiation or flare signals after amplitude modulation~\cite{Kiss2006,MorikawaNakamichi2023Entropy}.

Black-hole accretion systems give a magnetohydrodynamic extension.  Many local dynamo cells in a differentially rotating conductive flow may synchronize into mesoscopic magnetic structures.  Long X-ray light curves can then show Type A pink noise, while stochastic polarity flips and global reconnection may connect timing fluctuations to coronae, winds, and discrete jets.  The synchronized spin model formulates this idea explicitly by representing local accretion-flow dynamos as interacting macro-spins whose synchronization, excursion, and reversal organize both timing statistics and large-scale morphology~\cite{MorikawaNakamichi2026SSM}.  This line of interpretation is consistent with the broader view that $1/f$ noise, rms--flux relations, and lognormal variability in accreting systems reflect multiplicative, amplitude-modulated collective dynamics~\cite{Lyubarskii1997,UttleyMcHardyVaughan2005}.

\section{Discussion and conclusions}
\label{sec:discussion}
We have summarized a unified route to pink noise centered on economic indices.  The main point is that $1/f$ spectra can arise from a generic sequence: many interacting cycles $\rightarrow$ partial synchronization $\rightarrow$ beat and amplitude modulation $\rightarrow$ demodulation $\rightarrow$ observable low-frequency envelope.  This scenario naturally explains why economic data show two principal
empirical classes, Type A and Type B, together with mixed Type AB and
null Type O cases.
Extensive activity variables often display Type A/AB pink noise directly, whereas intensive price-like variables often display Type B pink noise only after detrending and demodulation.

The stochastic Kuramoto model is a minimal dynamical realization of this idea.  It represents repeated synchronization and desynchronization among many economic circulations and produces pink spectra over a broad critical domain.  It also gives a natural interpretation of the time-dependent PSD exponent as a coherence index.  Taylor's law and volatility clustering follow because the demodulated variance is itself a slowly fluctuating collective variable.

Several issues remain open.  First, the intrinsic demodulator must be identified in each economic observable.  Candidate mechanisms include transactions, price formation, inventory adjustment, bankruptcy, accounting, institutional thresholds, and nonlinear risk perception.  Second, the phase-only SKM should be extended to amplitude--phase dynamics to explain distribution functions, lognormality, power-law tails, and stronger Taylor-law scaling.  Third, apparent economic periods should be evaluated against modulated pink-noise null models before being interpreted as independent eigenperiods.  Finally, the same analysis should be applied systematically across economics, music, earthquakes, climate, stellar variability, solar activity, electric currents, and accretion systems.

The interpretation is limited in scope, but its central claims are falsifiable.
We do not claim that pink noise is a universal final explanation of economic fluctuations. Rather, we treat it as a diagnostic signature: when Type A, Type B, or Type AB \(1/f\) spectra appear, one should search for the synchronized units and for the demodulation process that exposes their slowly varying collective envelope. The scenario is supported if such units and demodulators can be identified, and it is weakened if the pink component can be explained without any coherent modulation or envelope-extraction mechanism.

\end{document}